\documentclass{webofc}
\usepackage[varg]{txfonts}   
\newcommand{\pp}{\ensuremath{\mathrm {p\kern-0.05em p}}}

\newcommand{\sqrtSnn}{\ensuremath{\sqrt{s_{\mathrm{NN}}}}}

\newcommand{\pt}{\ensuremath{p_{\mathrm{T}}}}

\newcommand{\MeVc}{\ensuremath{\mathrm{MeV}\kern-0.05em/\kern-0.02em c}}

\newcommand{\GeVc}{\ensuremath{\mathrm{GeV}\kern-0.05em/\kern-0.02em c}}

\newcommand{\GeVcSq}{\ensuremath{\mathrm{GeV}\kern-0.05em/\kern-0.02em c^2}}

\newcommand{\jpsi}{\ensuremath{{\rm J}\kern-0.02em/\kern-0.05em\psi}}

\newcommand{\dpt}{\ensuremath{\delta \kern-0.15em p_{\mathrm{T}}}}

\newcommand{\rpn}{\ensuremath{\Psi_n}}

\begin{document}
\title{Anisotropic flow of identified particles in Pb--Pb collisions at $\sqrt{s_{\rm NN}} = 5.02$ TeV}
%
%

\author{\firstname{Redmer Alexander} \lastname{Bertens}\inst{1}\fnsep\thanks{\email{rbertens@cern.ch}}}

\institute{University of Tennessee (Knoxville, USA)} 

\abstract{%
  Anisotropic flow is sensitive to the shear $(\eta/s)$ and bulk ($\zeta/s$) viscosity of the quark-gluon plasma created in heavy-ion collisions, as well as the initial state of such collisions and hadronization mechanisms. In these proceedings, elliptic ($v_2$) and higher harmonic ($v_3, v_4$) flow coefficients of $\pi^{\pm}$, $K^{\pm}$, p$(\overline{\rm{p}})$ and the $\phi$-meson, are presented for Pb–-Pb collisions at the highest-ever center-of-mass energy of $\sqrt{s_{\rm NN}}$ = 5.02 TeV. Comparisons to hydrodynamic calculations (IP-Glasma, MUSIC, UrQMD) are shown to constrain the initial conditions and viscosity of the medium. 
}

\maketitle

\section{Introduction}
Ultra-relativistic heavy-ion collisions are used to study the properties of the quark-gluon plasma (QGP), a state of deconfined quarks and gluons that is created at high energy densities and temperatures. Past measurements of azimuthal anisotropies in particle production relative to collision symmetry plane angles \rpn{} have shown that the QGP behaves as a nearly perfect fluid \cite{fluid}. These anisotropies in particle production arise when initial spatial anisotropies, resulting from the approximately elliptic overlap region of the colliding nuclei in non-central collisions, combined with the initial inhomogeneities of the system density, are transformed, through multiple interactions between the produced particles, into an anisotropy in momentum space. The efficiency of this process depends on the system's transport coefficients such as shear ($\eta/s$) and bulk ($\zeta/s$) viscosity.

Momentum anisotropy in particle production is quantified as harmonic coefficients $v_n$ of a Fourier series of the azimuthal angle ($\varphi$) distribution relative to the system's symmetry plane angles \rpn{} \cite{olli}
\begin{equation}\label{eq:flowdud}
    \frac{\mathrm{d}N}{\mathrm{d}\left(\varphi - \rpn\right)} \propto 1 + \sum_{n=1}^{\infty} 2 v_n \cos\left(n\left[\varphi - \rpn\right]\right).
\end{equation}
Harmonic coefficients $v_n$ -- commonly called \emph{flow coefficients} -- are, in addition to being a probe for $\eta/s$ and $\zeta/s$, also sensitive to the initial state of the system,  freeze-out conditions, hadronization mechanisms, and the lifetime of the system.

\section{Data analysis}
A sample of Pb--Pb collisions used for this work were recorded with the ALICE \cite{performance} detector at a center of mass energy per nucleon pair of \sqrtSnn{}~=~ 5.02~TeV. The data set comprises $\approx$ 6 $\times10^7$ collisions with a primary vertex reconstructed within $\pm$10 cm along the beam pipe. The Inner Tracking System (ITS) and Time Projection Chamber (TPC) are used to reconstruct charged particles at $\vert \eta \vert < 0.9$ and $\vert y \vert <$ 0.5. The V0 scintillator detectors, located at 2.8 $< \eta <$ 5.1 (V0A) and -3.7 $< \eta <$ -1.7 (V0C), are used for centrality determination, and reconstruction of the \textbf{Q}$_n^{\rm V0}$ vectors (see Eq.~\ref{eq:mth_sp}).  Particle identification is performed using ionization energy loss measured in the TPC, combined with the arrival time of particles in the Time of Flight (TOF) detector. The $\phi$-meson is reconstructed in the $\phi \rightarrow K^+ K^-$ channel, using the analysis method outlined in \cite{pidv2}.  

The scalar product method \cite{Voloshin:2008dg} is used to measure flow coefficients $v_n$, and is defined as
\begin{equation}
    v_{n}\{{\rm SP, V0C}\} = \langle \langle {\bf u}_n \cdotp {\bf Q}_n^{\rm V0C *} \rangle \rangle \Bigg/ \sqrt{ \frac{\langle {\bf Q}_n^{\rm V0C}\cdotp {\bf Q}_n^{\rm TPC *} \rangle \langle  {\bf Q}_n^{\rm V0C}\cdotp {\bf Q}_n^{\rm V0A *} \rangle} { \langle  {\bf Q}_n^{\rm TPC} \cdotp{\bf Q}_n^{\rm V0A *} \rangle } },
    \label{eq:mth_sp}
\end{equation}
where ${\bf u}_n=\exp(in\varphi)$, in which $\varphi$ represents azimuthal angle, is the unit vector of a particle of which $v_n$ is measured. Reference flow vectors \textbf{Q}$_n$ = $\sum_j \exp(i n \varphi_j)$ (the sum runs over all $j$ tracks and $^*$ denotes the complex conjugate) are measured in the TPC and V0 detectors, where V0C is chosen as reference flow detector as it still has a high event plane resolution. The large rapidity gap between ${\bf u}_n$ and ${\bf Q}_n^{\rm V0C *}$ ensures that contributions of short-range correlations that are unrelated to the initial geometry -- \emph{`non-flow'} -- are suppressed.. Brackets $\langle \dots \rangle$ in Eq.~\ref{eq:mth_sp} represent an all-event average; the double brackets in the numerator indicate that prior to the all-event average, an average over all  ${\bf u}_n \cdotp {\bf Q}_n^{\rm V0C *}$ within the single event is taken. 

\section{Results}

\begin{figure}
    \begin{center}
        \includegraphics[width=\textwidth]{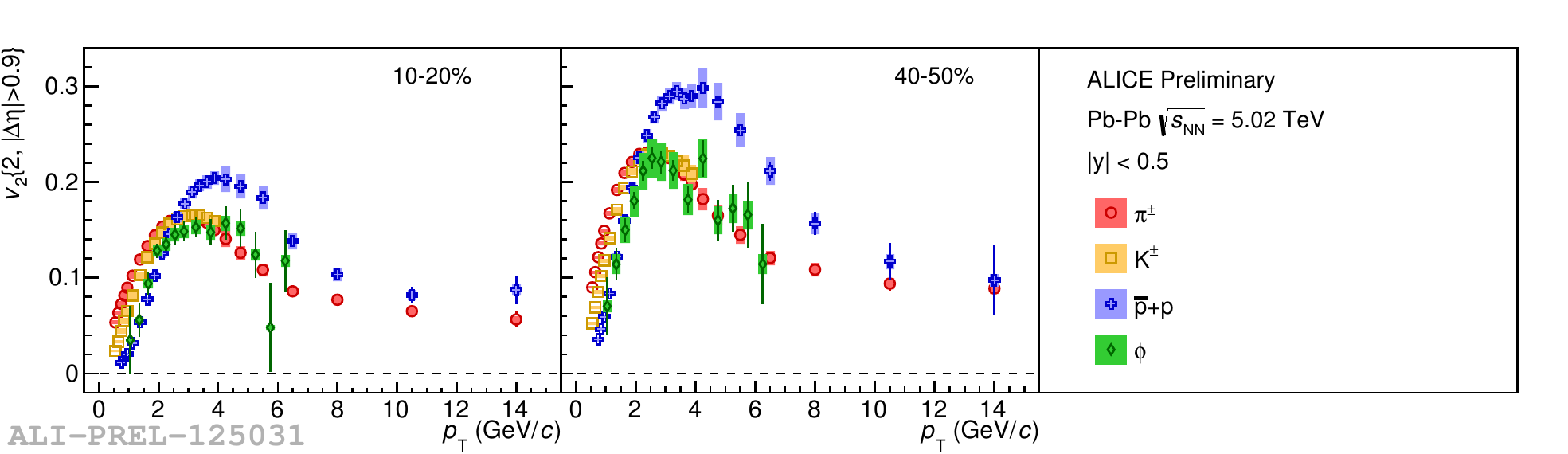}
        \includegraphics[width=\textwidth]{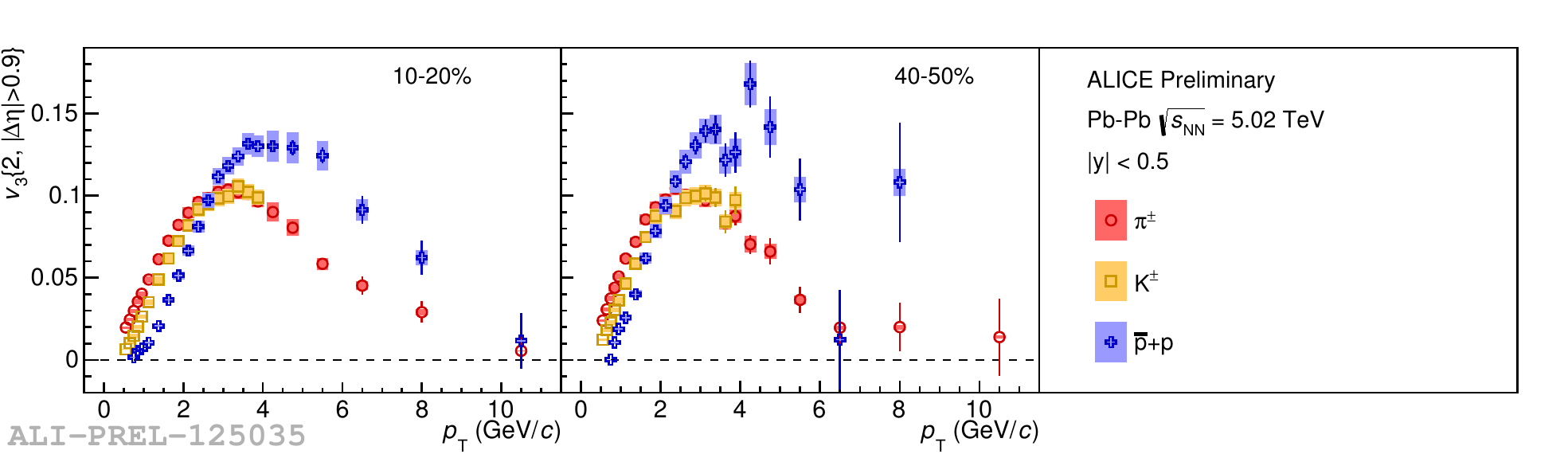}
        \includegraphics[width=\textwidth]{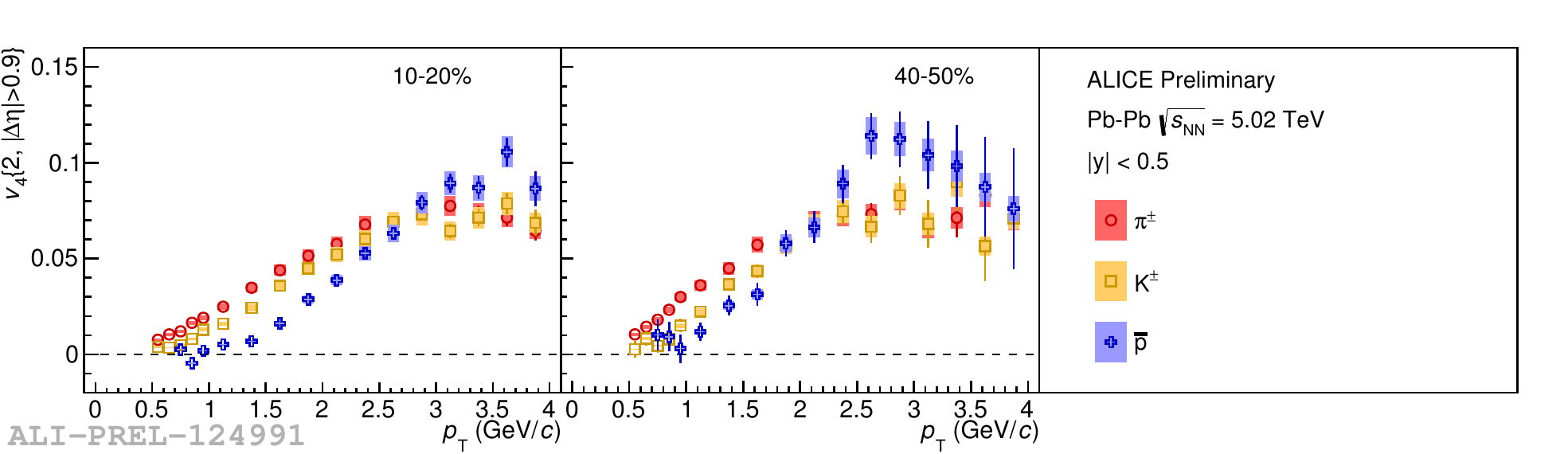}
        \caption{Flow coefficient $v_2$ (top), $v_3$ (middle) and $v_4$ (bottom) of $\pi^{\pm}, K^{\pm}$, p($\overline{\rm p})$ and the $\phi$-meson for 10-20\% (left) and 40-50\% (right) collision centrality as function of \pt{}. Statistical uncertainties are shown as bars and systematic uncertainties as boxes.}
            \label{fig:v2_pid}
        \end{center}
    \end{figure}

Figure~\ref{fig:v2_pid} shows \pt{}-differential $v_2$ (top panel) of $\pi^{\pm}, K^{\pm}$, p$(\overline{\rm{p}})$ and the $\phi$-meson for 10-20\% (left) and 40-50\% (right) collision centralities. For $\pt{} < 4~\GeVc{}$, only $\overline{\rm{p}}$ are considered to exclude a contamination from detector material. A mass splitting of $v_2$ of different particle species is found for \pt{}~$<$~2~GeV/$c$, which is indicative of strong radial flow \cite{radial}. For 3~$<$~\pt{}~$<$~8~\GeVc{} the curves for different particles group according to their constituent quark number, supporting the concept of particle production via quark coalescence \cite{dud2}. Flow of the $\phi$-meson most directly illustrates mass splitting and quark number scaling since the $\phi$ is a meson with a mass close to proton mass. The $\phi$-meson $v_2$ indeed follows proton $v_2$ at low \pt{}, but pion $v_2$ at intermediate \pt{}. The fact that p($\overline{\rm p}$) $v_2$ is larger than $\pi^{\pm}$ $v_2$ for $3 \lesssim \pt{} \lesssim$ 10 \GeVc{}, after which the $v_2$ converge, suggests that partonic energy loss is flavor independent at high \pt{}. 

    Inhomogeneities in the initial nucleon  distribution produce higher harmonic flow coefficients ($n > 2$). These coefficients are thought to be more sensitive to transport coefficients than $v_2$ \cite{dud3}. Non-zero $v_3$ is observed for $\pi^{\pm}, K^{\pm}$, p($\overline{\rm p})$ up to \pt{} $\approx$ 8 \GeVc{}, as is shown in the middle panel of Fig.~\ref{fig:v2_pid}; $v_4$ is non-zero in the entire measured range ($\pt{} < 4 \GeVc{}$, lower panel). The aforementioned mass splitting is visible for $v_3$ and $v_4$ up to $\pt{} \approx$ 2.5 \GeVc{}.

    \begin{figure}
        \begin{center}
            \includegraphics[width=\textwidth]{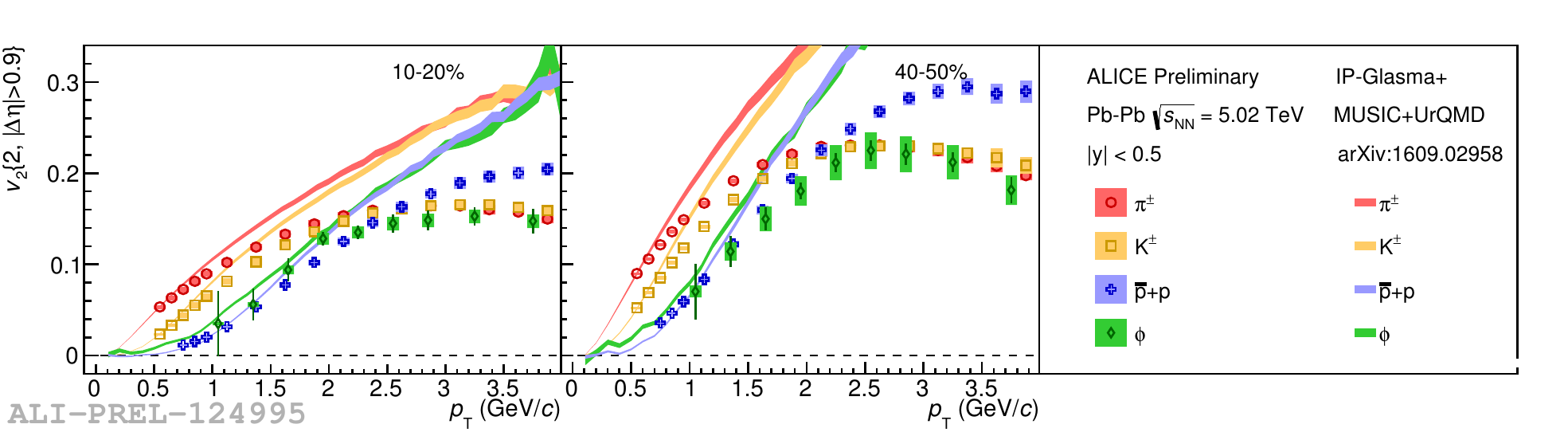}
            \caption{Flow coefficient $v_2$ of $\pi^{\pm}, K^{\pm},$ p$(\overline{\rm p})$ and the $\phi$-meson for 10-20\% (left) and 40-50\% (right) collision centrality compared to predictions from relativistic hydrodynamic calculations \cite{dud4}. Statistical uncertainties are shown as bars and systematic uncertainties as boxes.}
                \label{fig:v2_pid_hydro}
            \end{center}
        \end{figure}

        Model predictions from \cite{dud4} are shown to test the validity of the hydrodynamic description of the QGP in Fig.~\ref{fig:v2_pid_hydro}. The curves are based on an IP-Glasma initial state and use a viscous hydrodynamic medium evolution ($\eta/s$ = 0.095 with a temperature-dependent $\zeta/s$), followed by a hadronic cascade procedure for hadronization. Mass splitting is violated ($\phi$-meson $v_2 >$  p($\overline{\rm p}$) $v_2$) in the model, likely resulting from an under-prediction of the hadronic cross section of the $\phi$-meson. The model is in agreement with the data for $\pt{} <$ 1 \GeVc{} in central collisions, but overestimates $v_2$ already at lower \pt{} for more peripheral collisions.
        \begin{figure}
            \begin{center}
                \includegraphics[width=\textwidth]{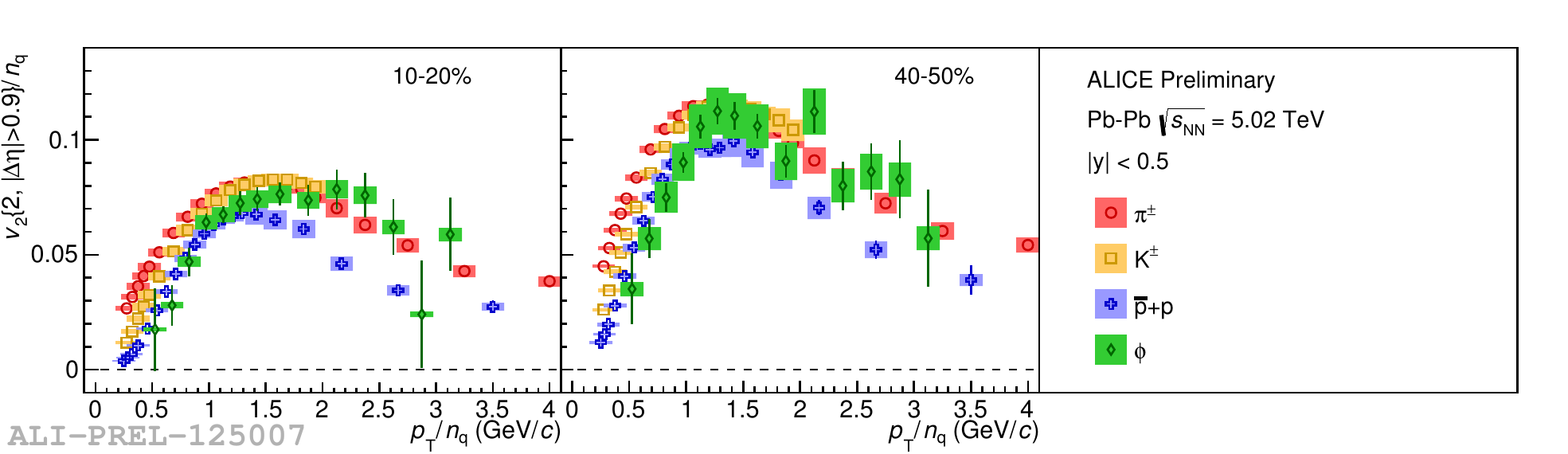}
                \caption{Scaling test for \pt{}-differential $v_2$ of $\pi^{\pm}, K^{\pm},$ p$(\overline{\rm p})$ and the $\phi$-meson for 10-20\% (left) and 40-50\% (right) collision centrality, The $x-$ and $y-$axes are scaled by the number of constituent quarks $n_{\rm q}$ for each species independently. Statistical uncertainties are shown as bars, systematic uncertainties as boxes.}
                    \label{fig:v2_pid_scaledPtNq}
                \end{center}
            \end{figure}
        
            To test the hypothesis of particle production via quark coalescence, the axes of Fig.~\ref{fig:v2_pid} are scaled by the number of constituent quarks $n_q$ independently for each species \cite{dud2} in Fig.~\ref{fig:v2_pid_scaledPtNq}. From \pt{}$/n_{\rm q} >$ 1.5 \GeVc{} particles group approximately according to their type (baryon, meson). Similar behavior is observed for $v_3$ and $v_4$ (not shown) but it is stressed that the observed scaling only holds approximately, as was also seen in \cite{pidv2}. 

            \section{Summary}
 Elliptic ($v_2$) and higher harmonic ($v_3, v_4$) flow coefficients of $\pi^{\pm}$, $K^{\pm}$, p$(\overline{\rm{p}})$ and the $\phi$-meson, measured in Pb–-Pb collisions at $\sqrt{s_{\rm NN}}$ = 5.02 TeV, have been measured. Mass splitting is observed for \pt{} $<$ 2 \GeVc{}, as well as approximate quark number scaling for \pt{} $>$ 2.5 \GeVc{}. These high-precision measurements can be used to put new constraints on model calculations, furthering the understanding of the initial state of heavy-ion collisions, as well as of the transport coefficients and lifetime of the QGP.

            \bibliography{references.bib}

\end{document}